\documentclass[sigconf,authorversion,nonacm]{acmart}

\usepackage{microtype}
\usepackage[normalem]{ulem}
\usepackage{ifthen}
\usepackage{xspace}
\usepackage{listings}
\usepackage{balance}

\usepackage{amsmath,amsfonts,amssymb}
\usepackage{lipsum,graphicx,multicol}
\usepackage{soul}
\usepackage{enumitem}

\newcommand{\eg}{\emph{e.g.,} }
\newcommand{\ie}{\emph{i.e.,}\xspace}

\newcommand{\myparagraph}[1]{\noindent\textbf{#1.}}
\newcommand{\system}{SOL\xspace} 
\newcommand{\cloudplatform}{Azure\xspace} 
\newcommand{\model}{\texttt{Model}\xspace}
\newcommand{\actuator}{\texttt{Actuator}\xspace}

\newcommand{\showComments}{false}
\newboolean{comments}
\setboolean{comments}{\showComments}
\ifthenelse{\boolean{comments}} {
    \newcommand{\yawen}[1]{{\textcolor{orange}{\textit[Yawen: #1]}}}
    \newcommand{\dan}[1]{{\textcolor{blue}{\textit[Dan: #1]}}}
    \newcommand{\rb}[1]{{\textcolor{green}{\textit[RB: #1]}}}
    \newcommand{\neeraja}[1]{{\textcolor{red}{\textit[neeraja: #1]}}}
    \newcommand{\christos}[1]{{\textcolor{brown}{\textit[christos: #1]}}}
} {
    \newcommand{\yawen}[1]{}
    \newcommand{\dan}[1]{}
    \newcommand{\rb}[1]{}
    \newcommand{\neeraja}[1]{}
    \newcommand{\christos}[1]{}
}

\newcommand{\tableref}[1]{Table~\ref{#1}}

\newcommand{\listref}[1]{Listing~\ref{#1}}

\newcommand{\secref}[1]{\S\ref{#1}}

\DeclareUnicodeCharacter{FB01}{fi}

\soulregister\eg7
\soulregister\ie7
\soulregister\system7
\soulregister\actuator7
\soulregister\texttt7
\soulregister\cite7
\soulregister\secref7
\soulregister\ref7
\soulregister\pageref7

\makeatletter
\AtBeginDocument{\let\hl\@firstofone}
\makeatother

\begin{document}

\title{SOL: Safe On-Node Learning in Cloud Platforms}

\newcommand*{\affaddr}[1]{\small{#1}} 
\newcommand*{\affmark}[1][*]{\textsuperscript{#1}}

\author{Yawen Wang}
\affiliation{%
  \institution{Stanford University}
  \country{Stanford, CA, USA}
}
\email{yawenw@stanford.edu}

\author{Daniel Crankshaw}
\affiliation{%
  \institution{Microsoft Research}
  \country{Redmond, WA, USA}
}
\email{dacranks@microsoft.com}

\author{Neeraja J. Yadwadkar}
\affiliation{%
  \institution{University of Texas at Austin}
  \country{Austin, TX, USA}
}
\email{neeraja@austin.utexas.edu}

\author{Daniel Berger}
\affiliation{%
  \institution{Microsoft Research}
  \country{Redmond, WA, USA}
}
\email{daberg@microsoft.com}

\author{Christos Kozyrakis}
\affiliation{%
  \institution{Stanford University}
  \country{Stanford, CA, USA}
}
\email{kozyraki@stanford.edu}

\author{Ricardo Bianchini}
\affiliation{%
  \institution{Microsoft Research}
  \country{Redmond, WA, USA}
}
\email{ricardob@microsoft.com}


\begin{abstract}
Cloud platforms run many software agents on each server node.
These agents manage all aspects of node operation, and in some cases
frequently collect data and make decisions.  Unfortunately, their
behavior is typically based on pre-defined static heuristics or offline
analysis; they do not leverage on-node machine learning (ML).
In this paper, we first characterize the spectrum of node agents in \cloudplatform
, and identify the classes of agents
that are most likely to benefit from on-node ML.  We then propose
\system, an extensible framework for designing ML-based agents that
are safe and robust to the range of failure conditions that occur
in production.  \system provides a simple API to agent developers and
manages the scheduling and running of the agent-specific functions
they write. We illustrate the use of \system by implementing three
ML-based agents that manage CPU cores, node power, and memory placement.
Our experiments show that (1) ML \hl{substantially} improves our agents, and (2) \system ensures that agents
operate safely under a variety of failure conditions.  We conclude
that ML-based agents show significant potential and that \system can
help build them.
\end{abstract}

\maketitle
\pagestyle{plain}

\thispagestyle{empty}

\section{Introduction}
\label{sec:intro}

\myparagraph{Motivation} 
Cloud platforms such as AWS, Azure, and GCP are complex.  In addition
to many control plane services running on dedicated capacity (\eg ~\cite{Borg,
  Protean,resourcecentral}), these
platforms run many management ``agents'' on each server node alongside
customer workloads.  The agents are responsible for configuring
and upgrading node software and firmware, creating and destroying
virtual machines (VMs), managing resource allocation and assignment
(\eg \cite{smartharvest, kumbhare2020predictionbased}),
checking for failure or vulnerability conditions (watchdogs),
monitoring resource health,
collecting telemetry, and many other tasks.

These tasks cannot be performed effectively from outside a node.
For example, resource
assignment must be fast (order of milliseconds) to prevent performance loss, power
management (\eg capping) must change hardware settings,
watchdogs need finer-grained telemetry than \hl{what} is unavailable off node, and
so on.
Because agents compete for resources
with customer workloads, platforms must
constrain their resource usage and/or at least partially offload
them to accelerator cards, as in \cite{nitro17, nitro19}.


%

Many agents collect data and make frequent decisions.
Currently, these decisions are based on static heuristics or
results of offline analysis.
But ML has shown potential to improve agent behavior
through workload- and hardware-aware decision-making~\cite{ML-Centric,hipster,smartharvest}.
For example,
an agent
responsible for conserving dynamic core energy can benefit from
learning the impact of core frequency on the workloads' performance at
each point in time.
Or a watchdog agent could learn to immediately flag serious issues with the
platform, while being slower for behaviors that are most likely benign.

Unfortunately, \hl{these} agents cannot take \hl{full} advantage of the ``centralized'' ML
systems currently available in production, such as
Resource Central~\cite{resourcecentral} or TFX~\cite{tfx}.
These systems train models offline (using data from all nodes) and serve
predictions (model inference) on-demand via a REST interface.
Thus, they are limited in their model update and inference frequency by the
network latency and bandwidth available for management communication.
In contrast, \hl{on-node} agents may need to operate on large amounts of fine-grained
node-local data (\eg core utilization samples collected every
tens of microseconds) and/or have to make high-frequency decisions (\eg
reassigning cores very few milliseconds)~\cite{smartharvest}.

\myparagraph{Challenges} We can overcome the limitations of 
centralized ML systems by
learning online on the nodes themselves.  However, 
deploying safe and robust learning on nodes that run (potentially
sensitive) customer
workloads poses challenges.  First, there are many conditions
that can lead ML-based agents to compromise quality of
service (QoS).  ML-based agents must be robust to unforeseen problems
due to the workload, the learning model, the node environment, or even
all three simultaneously, without needing human intervention or
communication with a centralized service.

Second, learning on individual nodes builds models online for cloud
workloads whose properties are unknown in advance.  This means that
the models themselves cannot be fully vetted offline ahead of time.
Thus, customer workloads must be protected from the decisions of bad
models running in production, rather than relying on a pre-deployment
process to filter out most of these inaccurate models before they ever
reach customers.  Possible causes for poor ML behavior include bad
input data due to corrupted or improperly configured hardware or OS
counters, or
attempting to learn from workloads that violate
modeling assumptions (such as steady-state or periodic workload
behavior).

Third, even accurate models can lead to sub-optimal agent behavior in
the presence of scheduling delays.
When the host resources are needed
for higher-priority tasks, such as virtual IO, the agent's execution
may be delayed and lead it to (1) miss important data samples and/or
(2) take actions based on stale data or a stale model.

Finally, customer workloads must be protected when the agent
experiences silent model failures or even hard crashes as a result of
external interference, unforeseen environmental conditions, or
software bugs.

\myparagraph{Our work} We first perform a comprehensive
characterization of the node agents that run in \cloudplatform
, and identify the classes of management
tasks that are most likely to benefit from on-node learning.
We find that three classes, which collectively make up 35\%
of all agents, can benefit from on-node learning.
Watchdog agents can use ML to both increase failure detection coverage
and detect problems earlier.
Monitoring agents can leverage ML
to adapt where and when they collect telemetry based on node activity, increasing
coverage without increasing cost.
Resource control agents can use ML to improve resource utilization while
protecting customer workload performance.

\yawen{clarifying objective/benefits/generality of SOL, and definition of "safety" in the following paragraphs}
\hl{When deploying on-node ML for these classes of agents, it is crucial to be robust to
the heterogeneous and evolving cloud environment under all failure conditions.
To facilitate the development and operation of robust on-node learning agents,
we introduce \system, a \textbf{S}afe \textbf{O}n-node \textbf{L}earning framework.}
\hl{Agent developers
  can} use \system{} for
developing ML-based agents that are internally safe and robust to the
range of failure conditions that can occur in production.
\hl{Different agents are typically developed by different teams in large cloud platforms.
  \system{} provides a unified interface across teams to reduce deployment complexity.
  Moreover, its interface allows cloud operators (\eg site reliability engineers or SREs) to safely terminate
  and cleanup after misbehaving agents without knowing anything about their implementation.}
\yawen{\sout{Although the failure conditions, required learning accuracy, and even
the very definition of accuracy (\eg when labels are available vs
when they are not) might differ across management tasks and agents,
the \emph{structure} of learning agents and the \emph{types of
problems} that can occur in practice are shared.}}

\hl{We design {\system} as a general framework to support a variety of on-node management agents that employ learning algorithms. By abstracting out structural similarity across learning agents and common types of problems the agents face in deployment, SOL presents a simple API with two key elements.} 
The first is a set of functions for developers to implement the four common operations for
ML-based control agents: collecting data, updating the model, getting a prediction from
the model, and actuating a change based on the prediction.
The second element is a set of required watchdog-style safeguards. The safeguards enumerate common failure conditions that can be hard to detect and debug in production. Agent developers must use the safeguards to internally
monitor different aspects of the agents, and avoid impact to customer QoS or node health when a problem is detected.
\hl{The safeguards ensure that agents are safe to deploy at scale alongside customer workloads.
  The exact definition of \emph{safety} varies based on the agents' purpose, but the desire to protect customer QoS and avoid wasting resources is common.
}


\system schedules and runs the developer-provided functions.
It also detects and informs the agent of any scheduling violations.
This is critical for avoiding the use of stale predictions under
highly dynamic workloads.

We demonstrate the use of \system by implementing three ML-based
agents. Each agent manages a different resource, uses a different
modeling approach, and has different data and scheduling constraints.
The first is a CPU overclocking
agent, SmartOverclock, that uses reinforcement learning to overclock workloads only
during the phases when they can benefit.
The second is a CPU-harvesting agent, SmartHarvest (introduced in~\cite{smartharvest}),
that predicts CPU utilization at
millisecond granularity to borrow idle cores and safely return
them before they are needed.
The third agent, SmartMemory, monitors each VM's memory usage to detect 
pages that can be migrated to remote memory without much performance impact.


\begin{table*}[t]
{\small
  \centering
  \caption{Taxonomy of production agents. The
  rightmost column lists whether the class could benefit from learning.}
  \vspace*{-.1in}
  \begin{tabular}{|l||c|l|l|c|}
    \hline
    \textbf{Class} & \textbf{Count} & \textbf{Description} & \textbf{Examples} & \textbf{Benefit?} \\
    \hline
    \hline
    Configuration & 25 & Configure node HW, SW, or data & Credentials, fire walls, OS updates & No\\
    \hline
    Services & 23 & Long-running node services & VM creation, live migration & No\\
    \hline
    Monitoring/logging & 18 & Monitoring and logging node's state & CPU and OS counters, network telemetry & Yes\\
    \hline
    Watchdogs & 7 & Watch for problems to alert/automitigate & Disk space, intrusions, HW errors & Yes\\
    \hline
    Resource control & 2 & Manage resource assignments & Power capping, memory management & Yes\\
    \hline
    Access & 2 & Allow operators access to nodes & Filesystem access & No\\
    \hline
  \end{tabular}
  \label{tab:node-agents}
}
\end{table*}

\dan{This paragraph might be part of the reason we got pushback from reviewers about our agents being novel, by claiming to demonstrate
that learning improves the agents which is not the focus of this paper.}
\myparagraph{Results}
\hl{We present a detailed experimental evaluation of our agents,
  first demonstrating that on-node learning significantly improves their
  efficacy, and then showing that agents implemented in {\system} operate safely
under a variety of failure conditions.}
As an example, SmartOverclock improves performance up to 41\% while
consuming 2.25x less power over a static overclocking baseline.
The \system safeguards limit the agent's power draw increase during
failure conditions to 18\%, while without the safeguards the same failure
condition leads to a 268\% power increase.

%

\myparagraph{Related work} We are unaware of similar agent
characterizations from commercial clouds.  The prior work on using
on-node learning has focused on ad-hoc resource management agents
(\eg \cite{smartharvest,hipster}), and did not consider
general frameworks for engineering safe ML-based agents.

\myparagraph{Contributions} In summary, our main contributions are:
\begin{itemize}[wide,labelwidth=!,labelindent=0pt,topsep=0pt,itemsep=-1ex,partopsep=1ex,parsep=1ex]
    \item A characterization of (1) the existing on-node agents in \cloudplatform 
    , and (2) the challenges involved in incorporating ML into them.
    \item The design of an on-node framework with an extensible API and runtime system for
      deploying ML-based agents that are robust to a wide variety of realistic issues.
    \item The implementation and detailed evaluation
      of three agents that demonstrate substantial improvements from on-node learning,
      while maintaining workload QoS and node health.
\end{itemize}

\section{Production On-Node Management}
\label{sec:hostagentstaxon}

\begin{table*}[t]
{\small
  \centering
  \caption{Examples of on-node learning resource control agents. Prior work has primarily focused on resource control.}
  \vspace*{-.1in}
  \begin{tabular}{|l||c|l|l|l|l|l|}
    \hline
    \textbf{Agent} & \textbf{Goal} & \textbf{Action} & \textbf{Frequency} & \textbf{Inputs} & \textbf{Model}\\
    \hline
    \hline
    SmartHarvest \cite{smartharvest} & Harvest idle cores & Core assignment & 25 ms & CPU usage & Cost-sensitive classification \\
    \hline
    Hipster \cite{hipster} & Reduce power draw & Core assignment \& frequency  & 1 s & App QoS and load & Reinforcement learning \\
    \hline
    LinnOS \cite{linnos} & Improve IO perf & IO request routing/rejection & Every IO & Latencies, queue sizes & Binary classification \\
    \hline
    ESP \cite{mishra2017esp} & Reduce interference & App scheduling & Every app & App run time, perf counters & Regularized regression \\
    \hline
    Overclocking~\secref{sec:scenarios} & Improve VM perf & CPU overclocking & 1s & Instructions per second & Reinforcement learning \\
    \hline
    Disaggregation~\secref{sec:scenarios} & Migrate pages & Warm/cold page ID & 100 ms & Page table scans & Multi-armed bandits \\ 
    \hline
  \end{tabular}
  \label{tab:use-case}
}
\end{table*}


Before delving deep into on-node learning and \system, it is important
to understand the spectrum of node agents in real cloud platforms, and
identify those that can benefit from learning.  Thus, in this section,
we first overview the classes of agents in \cloudplatform
and then discuss how learning can benefit a subset of the classes.






\myparagraph{Taxonomy of node agents} Regardless of whether an agent
runs on host CPUs or an offload card, it is typically a user-level
process responsible for a specific, narrowly-defined task.  This makes
agents simpler to develop, easier to maintain, less likely to impact node
performance, and less likely to affect each other in case of misbehavior.
It also makes them easier to categorize.  \tableref{tab:node-agents}
categorizes the agents in \cloudplatform into 6 classes.  There
are 77 agents, but many of them run rarely. Next, we provide an overview of
each class.

\noindent
{\em 1. Configuration} agents control aspects of the node's
hardware, software, and data.
They change the node state as directed by the platform's
control plane and run from every 10 minutes (configure TCP) to order of months (host OS upgrades).


\noindent
{\em 2. Service} agents run various node services that are critical for operating the cloud environment.
\hl{These services include VM lifecycle management, on-node agent creation, and security scanning and malware detection.}
They run throughout the lifetime of the node, at frequencies
ranging from seconds to minutes.


\noindent
{\em 3. Monitoring/logging} agents monitor and/or log data (off the node).
For logging fine-grained telemetry, they must aggregate/compress
data to reduce the amount to be sent off node.  They run at
different frequencies from the order of seconds to tens of minutes.



\noindent
{\em 4. Watchdog} agents (or simply watchdogs) check for problems that
either require telemetry that is only available on the node or where
detecting the problem off the node would be too slow to prevent
customer impact.
Watchdogs run fairly frequently, on the order of seconds to minutes.


\noindent
{\em 5. Resource control} agents dynamically manage resources, such as
CPUs, memory, and power.
Though they are not numerous, they run frequently, on the
order of seconds.


\noindent
{\em 6. Access} agents enable datacenter operators to diagnose and
mitigate incidents.
Some agents run continuously, while others only run when an incident
requires operator involvement.

\myparagraph{Runtime constraints on agents} Regardless of where they
run, agents compete for precious resources: on host CPUs, they compete
with customer workloads; on accelerator cards, they compete with other
agents and data plane operations.
Unconstrained agent execution may
introduce interference and tail latency effects~\cite{dean2013tail}.
For these reasons, each agent runs under strict compute and memory
constraints defined in its configuration (\eg 1\% CPU and 200MB of
memory for a host-CPU-based watchdog agent).  Agents also run at
lower priority than customer workloads and the host OS, which means
that agents may get temporarily starved or throttled.

\myparagraph{On-node learning opportunity} Today, production agents do
not take advantage of on-node learning and, thus, are not as effective
as they could be. \hl{Any agents that benefit from collecting data about current workload characteristics to guide dynamic adjustment of their behavior can potentially take advantage of ML.}\yawen{addressing reviewer's comment on design guidelines for on node learning agents}
In particular, we argue that resource control,
monitoring/logging, and watchdog agents can benefit significantly from
on-node learning.  

Resource control agents can benefit because the most efficient
assignment of resources (particularly compute, memory, and power) is
highly dependent on the current workload. Learning online directly on
each node can predict the short-term workload dynamics, and make
better assignment decisions without affecting customer QoS.  Better
assignments offer opportunities for improved efficiency and cost
savings. It is thus the focus of our case-studies
in~\secref{sec:scenarios}. Prior work has also considered
learning-based resource control agents (see \tableref{tab:use-case}).

Monitoring/logging agents can benefit because there is a cost to 
collecting samples and logging them off the node.  Yet these
agents, particularly those doing frequent sampling, treat every sample
as having the same value.
In steady-state, this results in oversampling, whereas in
highly-dynamic periods this can result in undersampling and the loss
of important information.
\hl{Online learning algorithms such as multi-armed bandits~\cite{bandits} can be used to}
smartly decide what telemetry to sample on the node or when to increase/decrease
sampling while staying within the collection and logging budget.

Finally, watchdogs can benefit because failure conditions can be
complex.  Existing watchdogs check for simple conditions that are
highly likely to indicate problems.  This often means that they cannot
detect problems until the problems are already affecting customer QoS.
Using on-node learning offers the opportunity to detect problems and
take mitigating actions earlier, as well as to detect and diagnose
more complex problems directly on the node.


\myparagraph{Summary, implications, and challenges} Production
platforms run numerous agents of different classes on each server
node. These on-node agents are resource-constrained and may be delayed by other activities.
We argue that resource control, monitoring/logging, and watchdog agents could
benefit substantially from on-node learning.  Agents from these
classes represent 35\% of the total node agents.
\hl{They run frequently and perform crucial or costly operations where ML can lead to significant savings (\eg improved resource utilization).}\yawen{addressing reviewer's comment on "35\% being a small fraction"}
{\em The key challenge is
  producing safe, robust, and effective ML-based agents under the
  constraints of real cloud platforms.}  Today, there are no systems
that can help agent designers address this challenge.

\section{On-Node Learning}
\label{sec:onnode}


%
%
%
%

On-node ML enables agents to become more agile and make smarter decisions, while considering fine-grained workload and resource utilization dynamics.
\tableref{tab:use-case} presents a selection of recent applications of ML to on-node management tasks \hl{that outperform static policies}.\yawen{addressing reviewers' comment on "ML vs static policies"}
These agents have different goals, employ different types of ML models, and learn from different telemetry.
\hl{Though these prior works demonstrate the benefits of incorporating ML logic into on-node agents, they neglect the impact of different failures conditions on agent performance and correctness, making them less practical to deploy into production.}


\subsection{When is on-node ML necessary?}


\myparagraph{Fresh workload-tuned models and predictions}
There are many workload dynamics that are only predictable a short window into the future.
For example, the SmartHarvest and the SmartOverclock agents rely on extremely short-term signals to predict
future CPU utilization (25ms into the future) and performance improvement from
overclocking (1s into the future), respectively.
For these use-cases, training on telemetry periodically logged to a centralized store would result
in perpetually stale models for the current workload.  Similarly, requesting predictions from centralized models to change a resource allocation
could result in perpetually predicting workload behavior that has already happened. Instead, the models must be trained and served online on the node, and continuously updated to learn the latest
behavior. \hl{This enables the agent to better respond to current workload behavior on the node.}

\myparagraph{Fine-grained telemetry}
For many of these use cases, the opportunities for improved efficiency come from learning high frequency
workload dynamics.
In these cases, the models must learn from telemetry sampled at a high enough rate to capture
these high-frequency effects.
For example, the SmartHarvest agent captures CPU telemetry every 50 $\mu s$
(the agent dedicates an otherwise idle core for capturing this telemetry;
when there are no idle cores, there is nothing to harvest so the agent
does not run), the SmartOverclock agent reads CPU counters every 100ms,
and the SmartMemory agent samples page access bits up to every 300ms.
This fine-grained telemetry cannot leave the node, as the size of the per-node data would
likely cause performance issues for customers.
For example, a single 16 GB VM whose memory is scanned every 300ms produces 100 MB of telemetry a minute.

\subsection{On-node ML challenges and requirements}
\label{sec:onnodechallenges}


%

\myparagraph{Bad input data}
On-node learning agents collect telemetry to update the model
and make decisions.
At the scale of a cloud platform operating millions of nodes, telemetry collection can fail in a variety
of ways \hl{(\eg misconfigured drivers, changes in data semantics between architecture or OS)}.\yawen{\sout{
For example, drivers can be misconfigured leading to invalid data readings or the semantics (\eg the units) of data
can vary between architecture or OS versions, or multiple processes reading and setting the same
counters can silently and transiently interfere with each other.}}

ML models are developed with implicit and explicit assumptions about data semantics, but will
often continue to learn and produce predictions on data that violates these assumptions.
The result is useless models trained on noise whose predictions should not be trusted.
Instead, data assumptions should be specified and explicitly checked.
In case of transient errors, if the invalid data can be detected and discarded,
the model can still learn and provide useful predictions.
Otherwise, learning on even small amounts of bad data can
corrupt the model.

\myparagraph{Poor model accuracy}
There are many reasons why a model may have poor accuracy.
For example, it may be learning from bad data (not all invalid data is detectable a priori).
Or it may be trying to learn the behavior of a workload that violates some modeling assumption
and is therefore unlearnable by this model
(\eg randomly changing workload dynamics).
Inaccurate models cause agents to take consistently bad actions, which can lead to
impact to customer workloads.
Instead, models must be evaluated continuously to ensure they meet accuracy expectations,
and their predictions should not be used during periods of poor accuracy.

\myparagraph{Unpredictable resource availability}
Agents are not guaranteed any computational resources.
During periods of high CPU demand on the host or expensive dataplane operations (\eg large amounts of virtual IO),
agents will be throttled for arbitrary periods of time.
Compute-intensive agents running close to their CPU limits can experience slowdowns resulting
in stale models and predictions.
As a result, the resource allocation or monitoring decisions made by an agent may be in effect for too long,
or the agent may make decisions based on stale data.
If not detected and handled appropriately, these delays can lead to \hl{unsafe agent behaviors (\eg negative impact on QoS)} by taking actions after workload dynamics have shifted.

\myparagraph{Node performance and reliability}
Finally, all agents must ensure that they are not negatively impacting node performance
or health in the face of opaque VMs.
Not all data or learning issues can be prevented, and other environmental factors outside the scope of the agent
may interfere with its operation (\eg VM live migration).
Sometimes servers run in stressed or constrained modes, such as being oversubscribed or power-capped.
In such settings, there can be little room for efficiency improvement from on-node learning and attempting
to do so can harm customer \hl{or node health}.
As a last line of defense, agents must estimate
their impact on client workloads and node health, and disable themselves if necessary.

\section{\system Interface and Design}
\label{sec:design}



\hl{We design {\system} to implement on-node ML agents that are safe to deploy alongside customer workloads by ensuring that they detect and mitigate all of the failure conditions from{~\secref{sec:onnodechallenges}}.}\yawen{clarifying the objective of \system in this paragraph}
\system's API guides agent developers through the agent-specific logic needed to manage these conditions, while
remaining highly extensible to different use cases. 
\hl{To implement a new agent in \system{},} all developers have to do is (1) write functions to instantiate the API's function signatures, and
(2) instantiate parameters for how often the functions need to run.
The \system runtime takes as input
the functions and parameters,
and manages scheduling and execution.


Next, we describe the interface \system exposes to agent developers, then discuss its runtime design and operation.

\subsection{\system interface}

\system is a lightweight C++ framework
that exposes an API to agent developers
which reflects the shared structure and failure modes of learning agents.
The API is split into two groups of functions: \model and \actuator, each with their own
sets of safeguards.
The \model is responsible for providing fresh and accurate predictions on a best-effort basis.
The \actuator makes control decisions at regular intervals (anywhere from milliseconds to minutes,
depending on the agent), using predictions from the \model when available.
The \model and \actuator run independently in separately scheduled loops so the
\actuator can continue to operate safely and take regular actions when the \model is
throttled or underperforming.


We adopt this split design to decouple the ML logic (\model) from the node management logic (\actuator), 
whether that is resource control,
monitoring decisions, or watchdog failure detection.
Enforcing strong abstraction boundaries simplifies agent design and ensures that the agent
is designed from the ground up to operate safely even without predictions.

\begin{figure}[t]
  \lstset{basicstyle=\small\ttfamily, keywords={interface, class}, frame=tb,
    label=lst:modelapi, captionpos=b, alsoletter={&},
    caption={\model interface. It is parameterized by the
    type \texttt{D} of the data collected and the type \texttt{P} of the prediction. Developers provide functions to instantiate these signatures.}
}
\begin{lstlisting}
interface Model<D,P>
{
  D CollectData();
  void UpdateModel();
  Prediction<P> ModelPredict();

  bool ValidateData(D data);
  void CommitData(Time time, D data);
  Prediction<P> DefaultPredict();
  bool AssessModel();
}
\end{lstlisting}
\vspace{-.15in}
\end{figure}

\myparagraph{\model interface} The top part of the \model interface (\listref{lst:modelapi}) specifies the three operations that all models take: (1) collect data to learn
and predict on, (2) update the model with newly acquired data, and (3) use the model to make predictions.
These three operations are called to form a single learning epoch.
Often models need to collect several datapoints before learning and making predictions, so a single learning
epoch can contain multiple data collection operations, but an epoch contains at most one model update and predict
operation.
The output of a successful learning epoch is a \texttt{Prediction} object that contains the predicted
value and an explicit expiration time for the prediction.
Data collection frequency, maximum duration, and the minimum and maximum number of data points that can be
collected in a learning epoch are all configurable by the developer.



The rest of the \model interface is devoted to detecting problems and taking mitigating action when they occur.\yawen{\sout{This increases the work of initially developing an agent, but ensures that it is appropriately hardened to common
operationalization problems that can impact customer QoS before the agent is deployed.
  Without this hardening, these problems can be difficult to detect and debug when they occur in production, hence reducing platform
  stability and leading to customer complaints.}}
Every individual datapoint must be validated,
and only if it successfully passes
validation will it be committed to \hl{be used in} the model.
\hl{The data validation interface in SOL takes as input the most recently read data. It can be used to perform range checks or simple distributional checks with developer-defined data structures in the class implementation.}\yawen{clarifying data validation in SOL}
There are limits to the extent individual telemetry data can be validated, 
\hl{but data validation helps ensure data points do not violate any testable properties of the inputs}.
In addition, developers must specify an \texttt{AssessModel} function that periodically checks whether
the model accuracy or other relevant performance metrics are acceptable \hl{for the prediction task of the agent.}
While the model assessment is failing, \system will
intercept predictions before they can be passed to the \actuator.
This means that the \actuator can assume that any predictions it receives are from a validated model.

For some agents, there are useful fallback heuristics that can be used to make safe workload-aware
decisions even without an accurate model.
These safe heuristic decisions can be implemented in the \texttt{DefaultPrediction} function.
\system will send them to
the \actuator instead when the model cannot produce an accurate prediction due to either data collection
or model quality issues.
\hl{\emph{Default} predictions should allow the node to behave in a way that has minimal impact on the agent-specific safety metric (\eg customer QoS), at the possible cost of running at lower efficiency.}\yawen{clarifying "impact of default prediction"}
However, even default predictions have an expiration time as they are still reliant
on fresh telemetry and can become stale.


Default predictions can also be explicitly sent to the \actuator at any stage of the learning epoch.
This short-circuits the current epoch and starts a new one.
This is useful when the developer can detect an error ahead of time (\eg if a prediction
is below a confidence threshold).

\begin{figure}[t]
  \lstset{basicstyle=\small\ttfamily, keywords={interface, class}, frame=tb,
    label=lst:actuatorapi, captionpos=b, alsoletter={&},
    caption={\actuator interface. This interface is parameterized by the type P of the prediction. Developers provide functions to instantiate these signatures.}
}
\begin{lstlisting}
interface Actuator<P>
{
  void TakeAction(Option<Prediction<P>> pred);
  bool AssessPerformance();
  void Mitigate();
  static void CleanUp();
}

\end{lstlisting}
\vspace{-.15in}
\end{figure}

\myparagraph{\actuator interface} By design, the \actuator interface (\listref{lst:actuatorapi}) closely resembles the interface of an agent that does not use ML.
It is a simple control function called \texttt{TakeAction} that is called either when new data becomes
available or after a developer-specified maximum wait time has elapsed, whichever comes first.
\hl{The only difference from non-learning agents is that learning agents use the prediction from a model}\yawen{\sout{as the data used}} to decide which action to take.

The \texttt{TakeAction} signature 
takes an \texttt{Option<Prediction>} type as an input.
There may not always be a prediction available from the model (even a default prediction), by the time
the \actuator must take an action, in which case the option type contains \texttt{None}.
Even if there is a prediction available, it may already be expired if there were scheduling delays
or throttling experienced by the agent.
\hl{{\system} detects scheduling delays by inserting various timestamp checks in the execution loop. It relies on the system clock for accurate timekeeping.}\yawen{clarifying how SOL detects scheduling delays}
 \hl{When a fresh prediction is not available within the specified time frame for the \actuator{} to take an action,} the agent should take a conservative, safe action to preserve customer QoS and node health, even if it comes
at the cost of reduced efficiency.


\hl{The \actuator{} requires its own safeguard specified in functions \texttt{AssessPerformance} and \texttt{Mitigate}.
\texttt{AssessPerformance} directly assesses the agent's behavior end-to-end,
independently of the internal state of the model. 
Whenever it detects the safeguard-triggering condition, the \texttt{Mitigate} function is then called to allow
the agent to take mitigating action.}
This safeguard serves as the last line of defense for the agent, and mirrors the existing approach 
in production, which requires node agents to have their own watchdogs.
\hl{The {\actuator} safeguard should measure proxies for the safety metric of the particular agent and define acceptable impact to these metrics as justified by business needs. For example, a poorly performing SmartHarvest agent can starve customer workloads that need CPU resources. Hence, its \texttt{AssessPerformance} function monitors vCPU wait time for these customer workloads and triggers the safeguard when the wait time exceeds a certain threshold (as configured by the developer). The \system{} runtime periodically evaluates \texttt{AssessPerformance} and calls \texttt{Mitigate} when the performance is unacceptable. The \texttt{Mitigate} function for SmartHarvest stops borrowing cores from customer VMs to stop the agent from impacting their performance.
However, failing the \actuator{} performance check is often a lagging indicator of negative impact. The safeguards in the \model{}
may detect and avoid problems before they trigger the \actuator{} safeguard, reducing the severity of impact.}\yawen{providing more insights on how to write safeguards and how safeguards are enforced in SOL}

Finally, the \actuator \yawen{\sout{also}} requires developers to provide an idempotent and stateless \texttt{CleanUp}
function.
This function can be safely called at any time (\eg by node SREs). \hl{It stops the agent and restores the node to a clean state},
regardless of whether the agent is running normally, has crashed, or is hanging.
\hl{SREs can also work with developers to define additional signals (\eg node health problems, frequent stalling of the agent) upon which agents should be cleanly terminated with the \texttt{CleanUp}
function.}
%

\subsection{\system runtime design and operation}

\myparagraph{Design principles} The key design decision in \system is to decouple the potentially expensive data collection and
ML component of the agent from the control component.
Internally, \system maintains two separate control loops running in separate threads.
The \model control loop collects data, updates the model, and produces predictions to a message queue.
The \actuator control loop consumes predictions from this queue when available and
periodically takes a control action and monitors the end-to-end scenario
performance.

The actuation logic is much simpler and less computationally expensive than the model logic, which may need
to collect substantial amounts of telemetry and perform many mathematical operations to train the model
or make predictions.
The specific actuation varies (\eg collect monitoring data, making resource control decisions,
trigger alerts), but the structure is the same.
At the same time, as we discuss in~\secref{sec:hostagentstaxon}, agents can run at best as
soft real-time systems that may be throttled or delayed without warning.

\yawen{\sout{Nevertheless, even under periods of heavy node activity, the agent is unlikely to be completely descheduled for long periods
of time.}}
By decoupling the expensive model logic from the lightweight actuation logic, we prevent the \model
from starving the \actuator during these periods of heavy throttling.
This provides an opportunity for the \actuator
to take a safe action to prevent node or customer impact while the model may be completely unable to run.

\begin{figure}[t]
  \lstset{basicstyle=\small\ttfamily, language=C++, keywords={interface, class}, frame=tb,
    label=lst:solruntime, captionpos=b, alsoletter={&},
    caption={Executing an agent. Once developers have implemented the \system interface,
    they pass their implementation to the \system runtime for scheduling and execution.}
}
\begin{lstlisting}
class Schedule
{
  // Model
  int data_per_epoch;
  duration data_collect_interval;
  duration max_epoch_time;
  duration assess_model_interval;
  // Actuator
  duration max_actuation_delay;
  duration assess_actuator_interval;
}
void main()
{
  Schedule schedule(config_file);
  Model* model = new OverclockModel();
  Actuator* act = new OverclockActuator();
  SOL::RunAgent(model, act, schedule);
}
\end{lstlisting}
\vspace{-.15in}
\end{figure}

\myparagraph{Operation} Given an instantiation of the agent API, \system automatically starts and runs the \model and \actuator
control loops according to developer-provided schedules (\listref{lst:solruntime}).
The \model loop collects data at the frequency specified by the user until either enough data has been collected
and validated or the maximum epoch time has elapsed.
If enough data has been collected, \system updates the model and makes a prediction.
Otherwise, it short-circuits the learning epoch by sending a default prediction to the \actuator.

In addition, \system assesses the model accuracy periodically (every K epochs as specified by the user).
If the model fails the accuracy check, \system continues to operate the \model control loop normally.
However, \system intercepts predictions and instead passes a default prediction to the \actuator. This still allows the model to be updated and produce predictions, hence providing the opportunity
for the model to recover from a period of bad performance.
\hl{At the same time, it prevents the \actuator{} from acting on bad predictions.}

The \actuator waits on the prediction message queue for up to a maximum wait time.
When new predictions are available, it immediately uses them to take actions.
If a timeout occurs, \system still calls \texttt{TakeAction} to provide
an upper bound on the time between control actions in the agent.
\system also periodically checks the \actuator safeguard to detect behavior that could impact customer
workloads or node health.
If the safeguard is triggered, it
halts the \actuator control loop until
the unsafe behavior is no longer detected.

\section{Developing Agents in \system}
\label{sec:scenarios}


\yawen{adding a paragraph on supporting new agents in SOL and the development burden}
\hl{To build new agents in \system, developers need to provide the implementation of the four common ML operations along with the various safeguards.
{\system}'s APIs direct development efforts towards handling failure conditions ahead of time via the definition of safeguards. This requires developers to carefully reason through what conditions are appropriate to monitor and what mitigating actions should be taken in response. We argue this extra development burden up front is crucial, as it helps substantially reduce the complexity of managing learning agents in production. The operationalization complexity has been shown to contribute a significant part of the total cost for deploying ML in production~\cite{sculley2014machine}.}

We implemented three on-node learning agents using \system.
\hl{They differ in the type of node resources they manage, the input data and ML models they use, and the timescales they run at.}  
\hl{Next}, we discuss how these agents can benefit from learning,
and their implementation in \system.
\hl{Unless otherwise stated, the various agent parameter values were selected based on experimental tuning.} 
In~\secref{sec:eval}, we demonstrate the consequences of running these agents unchecked during failures
and how \system minimizes these consequences by detecting and mitigating when failures occur.

\subsection{CPU overclocking}
\label{sec:subsec:oc-agent}

CPU overclocking presents opportunities for substantial performance improvements on some
workloads~\cite{jalili2021cost-ef}.
However,
overclocking significantly increases power consumption
and can shorten hardware lifetimes.
As cloud platforms explore providing overclockable VM offerings, they want
to balance the performance improvements with the extra power cost.

To address this problem, we created an intelligent on-node overclocking agent called SmartOverclock, which uses Q-learning~\cite{sutton2018reinforcement}, 
a simple form of Reinforcement Learning (RL).
It monitors the average Instructions Per Second (IPS) performance counter across the cores of each VM and learns when to overclock the VM.
At the end of every 1-second learning epoch, the agent uses the observed IPS and current core frequency
to calculate the current RL state and reward.
It then updates the RL policy and uses it to pick the frequency
for the next learning epoch.
Because the agent cannot directly observe workload-level metrics (\eg tail latency) inside opaque VMs, it assumes that a workload
benefits from overclocking when higher CPU frequencies increase IPS.
\hl{Though IPS is not a perfect proxy for identifying whether overclocking is beneficial, it works well for most optimized workloads.} 
To balance exploitation of the policy learned so far with exploration of new frequencies,
the agent uses the action selected by the RL policy 90\% of the time and randomly
picks a frequency 10\% of the time.

\myparagraph{Validating data}
The agent collects multiple CPU counters and validates that 
they are within their expected ranges, discarding any
data that fails this check, \hl{\eg the IPS value should be between 0 and}
$\texttt{max\_freq}*\texttt{max\_IPC}$.
Even a small fraction of bad data can cause the
model to learn a sub-optimal policy and prevent
workloads from benefitting from overclocking (see~\secref{sec:eval}).


\myparagraph{Assessing the model}
A poorly performing RL policy can cause the agent to overclock workloads that do
not benefit, resulting in wasted power.
To detect a bad policy, the agent (in the \texttt{AssessModel} function)
computes the difference, $\Delta_{r}$, between the \emph{observed} reward when overclocking
and the \emph{expected} reward from using the nominal frequency. 
It discards predictions if the average $\Delta_{r}$ over the last 10 epochs falls below a threshold.
In this case, the agent continues to randomly explore, but overrides the RL-selected
actions by always picking the nominal frequency as the default prediction.

%

%

\myparagraph{Handling delayed predictions}
\hl{Brief periods of wasted power are acceptable for the agent to recover from short scheduling delays. Thus,} the Actuator will wait for up to 5 seconds \hl{(5 learning epochs)} for a prediction.
If it has not received an un-expired prediction at the end of this period, it takes the safe
default action of setting the CPUs to the nominal frequency to avoid wasting power.

\myparagraph{Safeguarding the Actuator}
As the end-to-end safeguard for the Actuator, we define a factor
$\alpha$, using three CPU counters: $\alpha = (\textnormal{unhalted\_cycles - stalled\_cycles})/{\textnormal{total\_cycles}}$.
This factor
serves as a binary indicator of whether a workload might benefit from overclocking.
If $\alpha$ is low,
the workload will not
benefit much and overclocking would simply waste power.
The Actuator monitors the 90$^{th}$-percentile (P90) of $\alpha$ values over the past 100 seconds
and triggers the safeguard if this value is below a threshold.
\hl{We use P90 to smooth transient drops in $\alpha$, while quickly exiting the safeguard when activity increases again.}
\yawen{\sout{We use the P90 value to only trigger the safeguard during
stable periods of low activity, and quickly exit the safeguard when activity increases again.}}
The safeguard restores all cores to the nominal frequency in the \texttt{Mitigate} function.

%
%

\myparagraph{Cleaning up}
The \texttt{Cleanup} function kills any running SmartOverclock agents
and then restores all cores to the nominal frequency.

\subsection{CPU harvesting}
\label{sec:subsec:smartharvest-agent}

The second agent we implement in \system is the SmartHarvest agent from \hl{prior work}~\cite{smartharvest}.
\hl{We adopt the same model design and parameters values as used in ~\cite{smartharvest}.}
This agent opportunistically ``harvests'' CPU cores that have been allocated to a (primary) set of VMs but are currently 
idle. 
It then loans the harvested cores to a special VM (called an ElasticVM), but must return the cores
to the primary VMs as soon as they need them.
\hl{Prior work has shown that ML is beneficial for this task~\cite{smartharvest}, but it did not fully explore the design of safeguards to ensure safe and robust agent performance.
  We choose it as a case study to demonstrate the benefits of implementing previously explored ML use cases in \system{} (see{~\secref{sec:eval}}).} 

The agent uses a cost-sensitive classifier from the VowpalWabbit framework~\cite{vw} to predict the maximum number
of CPU cores needed by the primary VMs in the next 25 ms.
It collects VM CPU usage data from the hypervisor every 50$\mu s$ and computes distributional features
over this data as input to the model.


\myparagraph{Validating data}
We perform range checks on the counter readings similar to those of the SmartOverclock agent.
In addition, if the primary VMs use all their allocated cores during a learning epoch, it is impossible to
distinguish whether they needed exactly that many cores, or whether they were under-provisioned
during the epoch and the degree of that under-provisioning.
Learning from this CPU telemetry can skew the model and cause it to systematically underpredict primary
core usage.
We therefore also discard any data collected during periods of full utilization by the primary VMs, as done in~\cite{smartharvest}.

\myparagraph{Assessing the model}
The original SmartHarvest designers did not discuss approaches to assessing the accuracy of the learning model, instead relying on their
version of the Actuator safeguard to detect and mitigate any problems.
However, the Actuator safeguard, while important, is a lagging indicator of impacted
performance.
Assessing the model accuracy can detect some problems earlier.
Thus, our implementation measures the percentage of time that predictions from the model lead to primary VMs running out of idle cores.
If this percentage is high, the model safeguard is triggered.

\myparagraph{Handling delayed predictions}
Similar to SmartOverclock, our implementation of SmartHarvest sets a time limit on the wait
time for a prediction produced
by the model.
The agent waits for a maximum of 100 ms \hl{(4 learning epochs)} to account for
its tighter harvesting control loop.

\myparagraph{Safeguarding the Actuator}
The agent uses a hypervisor counter reflecting how long virtual cores of a primary VM have waited for physical cores to run on as a proxy for workload QoS degradation.
Long wait times indicate insufficient idle cores.
The Actuator safeguard monitors the P99 wait time using the same approach as in~\cite{smartharvest}. If the value is high, it disables harvesting by giving all cores back to the primary VMs.

\myparagraph{Cleaning up}
\texttt{Cleanup} kills any running SmartHarvest agents
then returns all harvested cores to the primary VMs.

\subsection{Page classification for tiered systems}
\label{sec:subsec:coldmemory-agent}

Our third SOL-based agent, called SmartMemory, targets managed two-tiered memory systems, where a slower and lower-cost byte-addressable memory (\eg persistent~\cite{yan2019nimble,kim2021exploring} or disaggregated memory) sits behind the faster but expensive DRAM-based first tier.
%
To efficiently use such systems, prior work exploited the highly-skewed popularity of
pages in real-world workloads~\cite{yan2019nimble,kim2021exploring}.
Building on this idea, our agent seeks to identify pages as hot, warm, and cold, so that a small number of hot pages
are stored in first-tier DRAM~\cite{yan2019nimble,kim2021exploring},
warm pages are on slow memory, and cold pages are compressed or not stored at all~\cite{lagar2019software}.

To determine page hotness, we can scan page access bits through the hypervisor~\cite{park2019profiling,yan2019nimble,kim2021exploring}.
Frequent scanning provides more fine-grained information about relative page access rates, but may also degrade workload performance from TLB misses.
Each time a page's access bit is cleared, the page entry is flushed from the TLB.


Our agent uses ML to minimize
the number of TLB flushes while still accurately classifying memory as hot/warm/cold.
It uses Thompson Sampling~\cite{thompson1933,thompsonsamplingfnt} with a Beta distribution prior,
a well-known multi-armed bandit~\cite{bandits} algorithm that yields good performance in practice.
The agent learns the best scanning frequency for each 2MB region of memory, divided into 512 4KB pages.
The optimal scanning frequency is the lowest frequency that yields the same number of
accesses as the maximum frequency.
This forces hot batches to be sampled at the maximum frequency, while colder batches can be sampled much less often.
In every epoch, the agent uses the Thompson Sampling models to decide how often to scan each batch,
ranging from 300ms to 9.6s.
At the end of each 38.4-second epoch (4x the maximum sampling period of 9.6s), the agent observes
whether each batch was oversampled, undersampled (as approximated by
number of consecutive access bits set), or well sampled, and updates the models accordingly.
The model then uses the variable rate scans to estimate the minimal set of batches
that contributed 80\% of total memory accesses.
It classifies these batches as hot, and the remainder as warm batches that are candidates
for first-tier DRAM offloading.
Similar to \hl{the heuristic used in previous work on cold memory detection}~\cite{lagar2019software},
we treat batches that have been untouched for more than 3 minutes as cold and exclude
them from scanning and our analysis.

%
\myparagraph{Validating data}
The access scanning driver will return an error code if it fails
to scan or reset any access bits.
In these cases, \texttt{ValidateData} fails the sample.


\myparagraph{Assessing the models}
The main risk from inaccurate models is that hot memory regions will be undersampled,
leading the agent to conclude they are colder than they really are.
The SmartMemory model randomly samples 10\% of the batches at the maximum frequency and computes
the total number of accesses to these batches.
It uses this sample as ground truth to estimate the fraction of access bits missed by
the model-recommended scanning rates.
If the fraction of missed accesses rises above 25\%, the model is deemed to be undersampling
page accesses.

%

To provide safe default predictions under partial sampling or undersampling,
the agent downsamples the access scans from all the batches to the lowest scanning frequency
so that hit counts across different batches are directly comparable.
It then
targets a much more conservative 95\% hottest batches to
keep in first-tier DRAM, selecting only the coldest 5\% of batches as candidates for warm memory using
these downsampled hit counts.
This helps protect workload QoS without completely disabling the second tier.

\myparagraph{Handling stale predictions}
Unlike our other agents, SmartMemory has no need to take any immediate mitigating action
when predictions are delayed.
It simply leaves the hot and warm pages where they are.
\hl{If this decision becomes stale before the next prediction is received,
the non-blocking system design triggers the Actuator safeguard 
to mitigate the problem.}

\myparagraph{Safeguarding the Actuator}
The agent can directly observe the number of memory accesses to each tier
using existing hardware counters.
If the fraction of remote accesses over the last epoch is
above the 20\% target service level objective (SLO),
the Actuator safeguard is triggered.
In this case, the Actuator immediately migrates the 100 hottest batches
in the second-tier memory back to the first tier.
If the first tier does not have room for all 100 batches, it migrates as many as possible
starting with the hottest batch.

\myparagraph{Cleaning up}
\texttt{Cleanup} kills any active SmartMemory agents
and restores all second-tier batches back to the first tier
until either all batches have been restored or the first tier is full.




\section{Evaluation}
\label{sec:eval}

We evaluate (1) the utility of ML in each \system agent we build,
and (2) the efficacy of \system's API in detecting failure conditions and mitigating
their impact.
Each agent we study manages a different resource.
Therefore, the impact of failures on workload performance and node health also differs across agents.
We begin by evaluating all safeguards using the
SmartOverclock agent
and include additional experiments for the SmartHarvest and
SmartMemory safeguards where their behavior differs significantly
from SmartOverclock. 

\subsection{Experimental Setup}

\hl{Because agents running on each node are independent of each other, we run experiments on a single node and inject failures into the system to evaluate their resilience to these failures.} \yawen{clarifying experiment setup}

All experiments run on a two-socket Intel server with the Xeon
Platimum 8171M processor capable of running at up to 2.6GHz, with 26 cores per socket
and
384 GB DRAM. 
To reduce performance jitter for the customer VMs, we disable simultaneous multithreading, C-states, and Turbo-Boost. 
The server runs the Hyper-V hypervisor. 


We run the agents in user-space on the root partition of Hyper-V. \hl{The overhead of running an agent is dependent on invocation frequency and computation overhead of various learning functions and safeguard checks. The \system{} runtime manages the scheduling in user-space and runs in the same process as the agent, requiring very few resources.} \yawen{clarifying overhead of SOL}


\begin{figure}[t]
    \includegraphics[width=\linewidth]{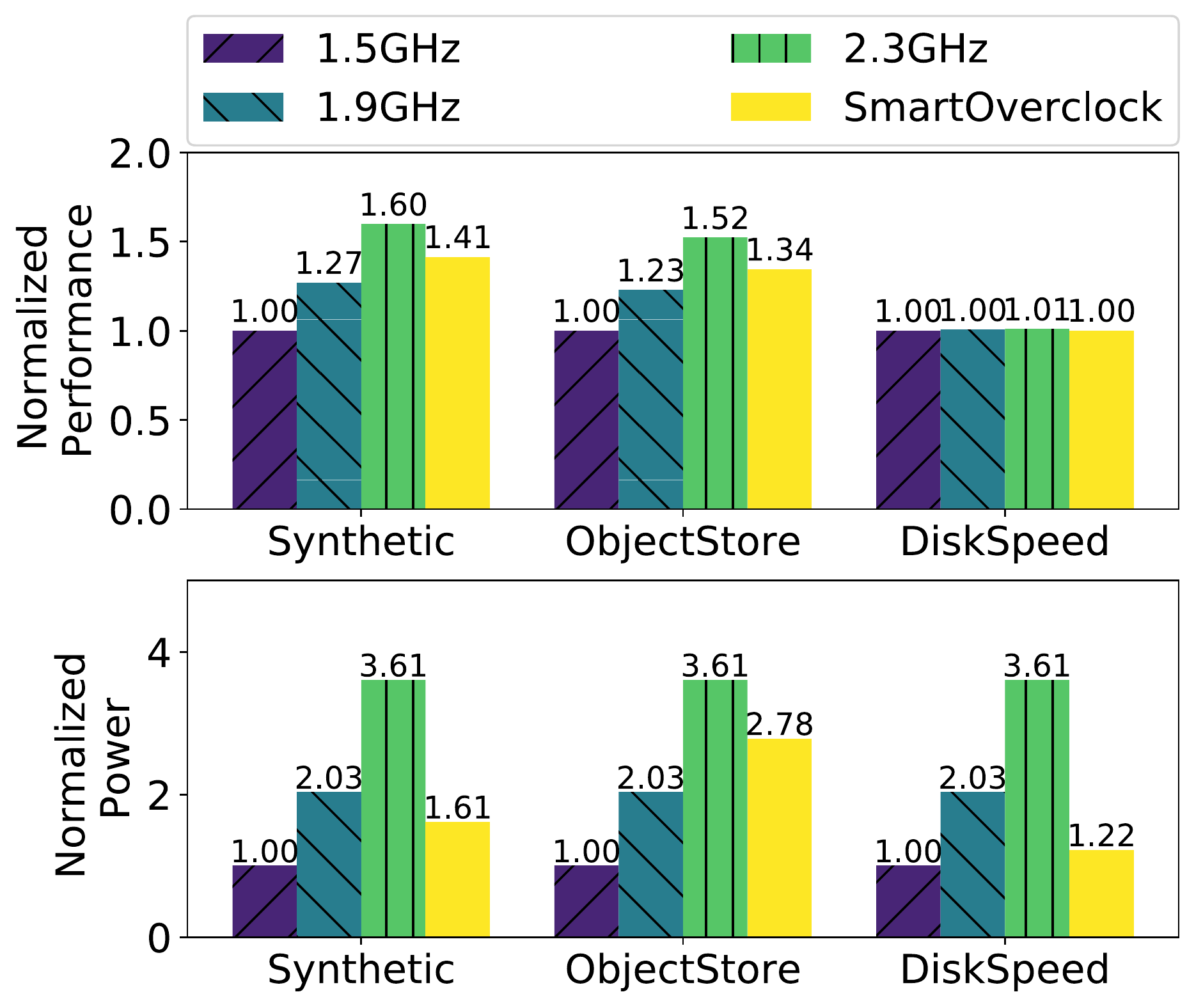}\par
    \vspace{-.1in}
    \caption{SmartOverclock learns to only overclock when the workload can benefit.
    Performance and power are normalized to the baseline values at 1.5GHz.}
    \vspace{-.15in}
    \label{fig:smart_overclock}
\end{figure}

\begin{figure*}[t]
\begin{multicols}{2}
    \includegraphics[width=0.7\linewidth]{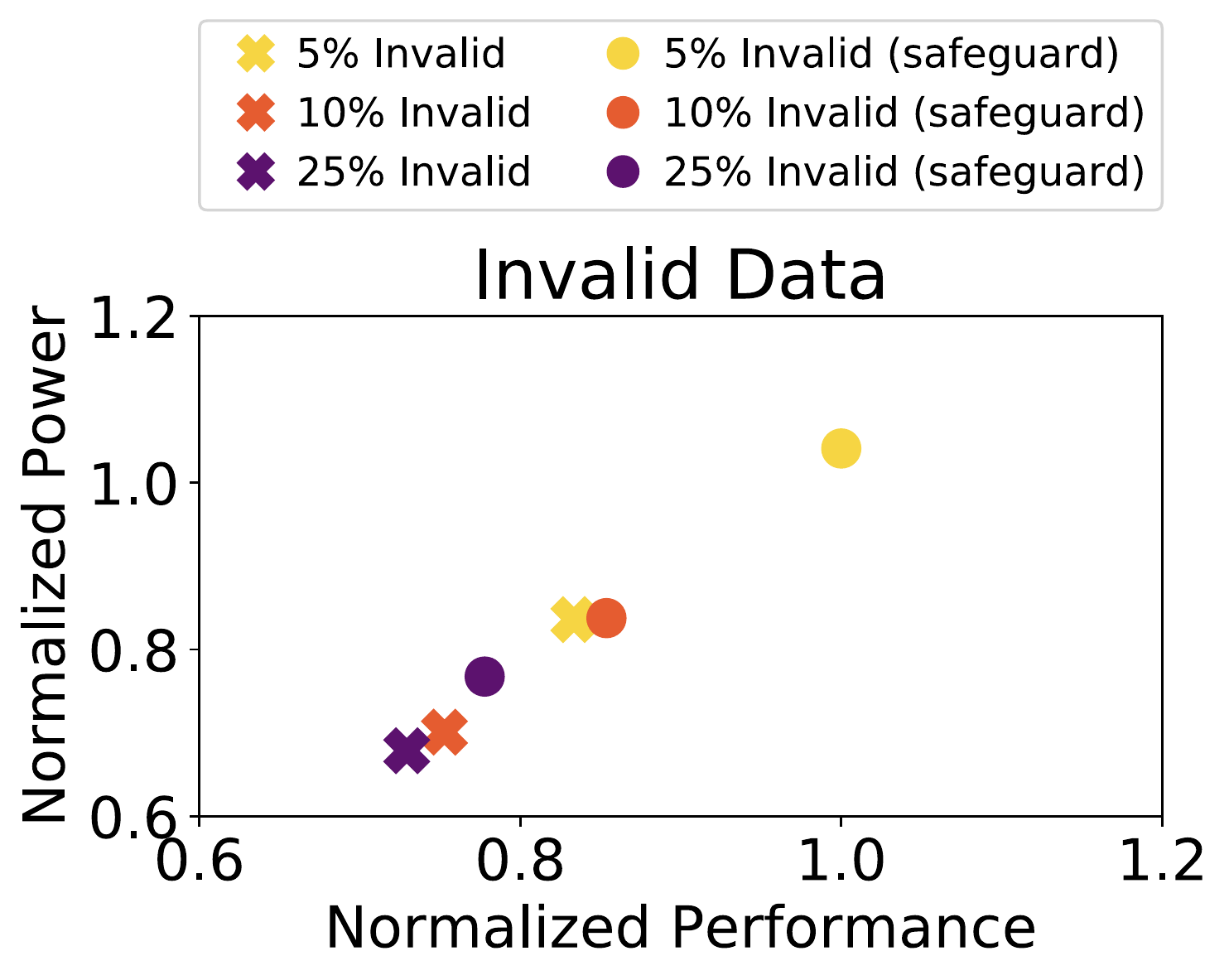}\par
    \vspace{-.1in}
    \caption{SmartOverclock data validation safeguard mitigates transient data errors.
    Power and performance are normalized to the ideal agent decision-making (all valid data).}
    \label{fig:oc:data_safeguard}

    \includegraphics[width=0.7\linewidth]{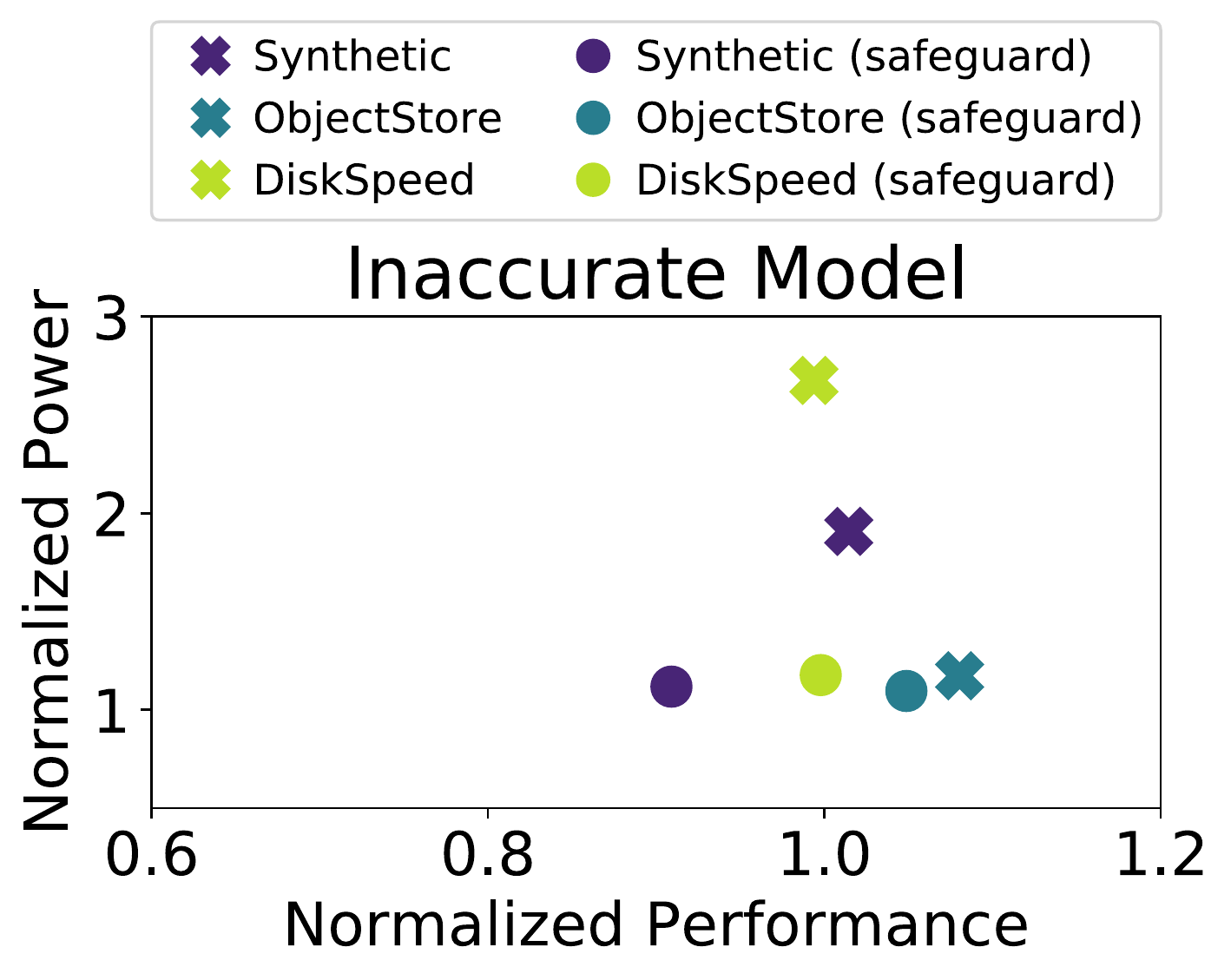}\par
    \vspace{-.1in}
    \caption{SmartOverclock model safeguard detects when RL overclocks without gains.
     Power and performance are normalized to the ideal agent decision-making (correct model).}
    \label{fig:model_safeguard_wrong_model}
\end{multicols}
\vspace*{-.2in}
\end{figure*}

\begin{figure*}[t]
\begin{multicols}{2}
    \includegraphics[width=\linewidth]{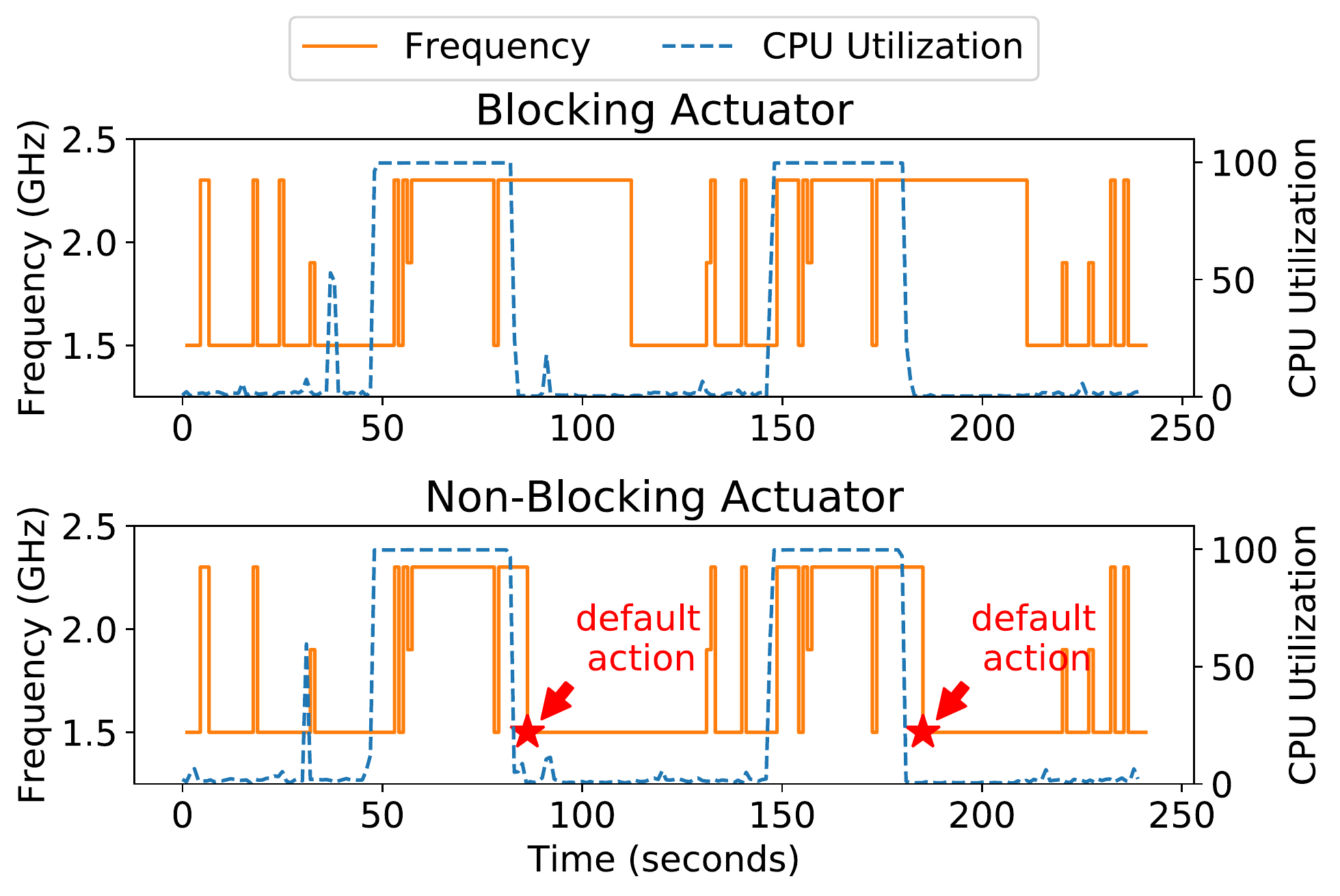}\par
    \vspace{-.1in}
    \caption{Non-blocking Actuator for SmartOverclock prevents wasted power when predictions are unavailable.
    }
    \label{fig:scheduling_safeguard}

    \includegraphics[width=\linewidth]{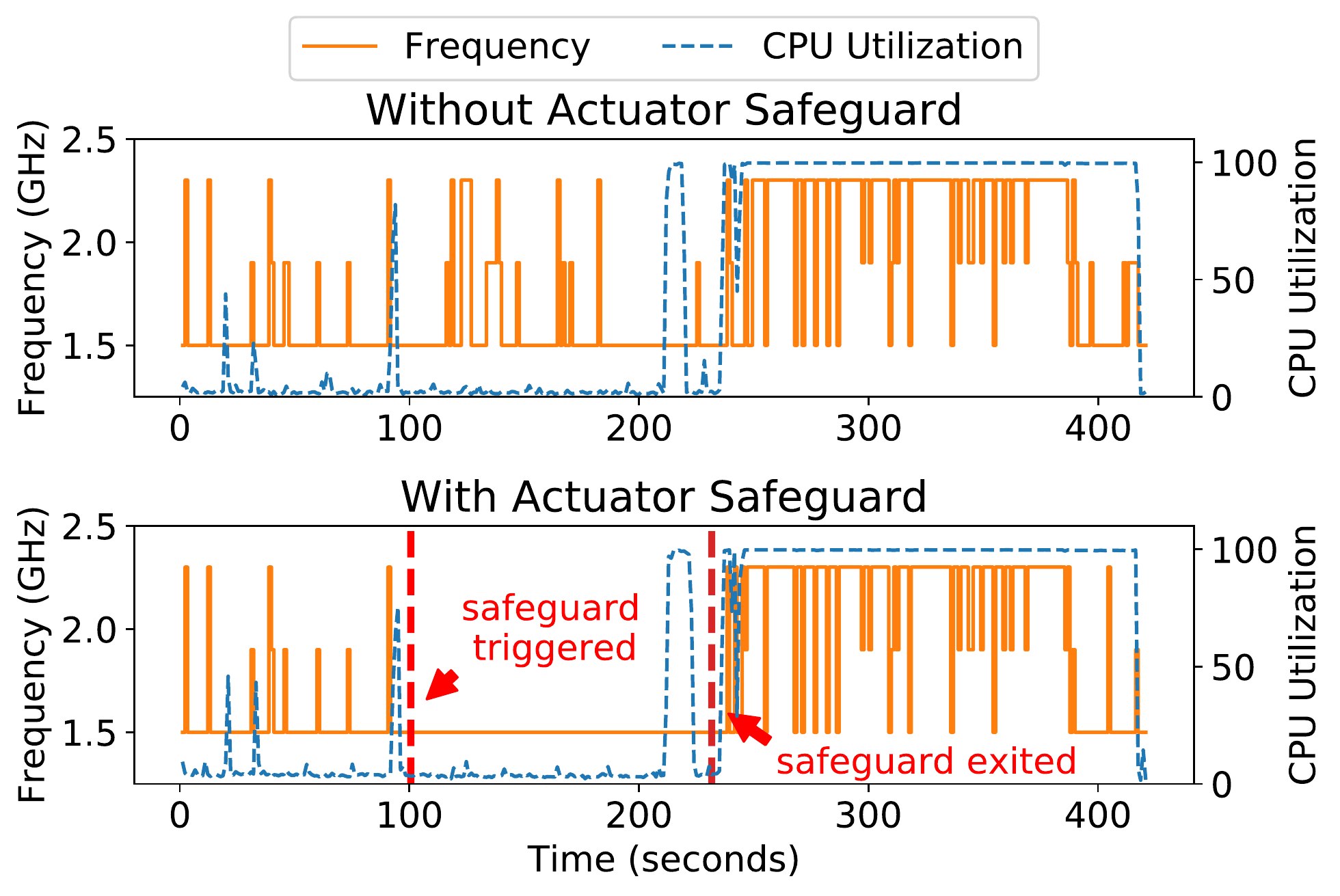}\par
    \vspace{-.1in}
    \caption{The SmartOverclock actuator safeguard reduces wasted power during long-lasting idle phases.}
    \label{fig:scenario_safeguard}
\end{multicols}
\vspace*{-.15in}
\end{figure*}

\subsection{SmartOverclock}

We set the nominal server frequency to 1.5GHz 
and let the agent select from three possible CPU frequencies: 1.5, 1.9, and 2.3 GHz.
Within an epoch, the SmartOverclock agent sets cores for a VM to the
same frequency, but can change this frequency between epochs.
While per-core frequency scaling is possible, the agent has no
visibility into the thread
scheduling (which governs per-core utilization) within the VM.


We first compare workload performance and power consumption of SmartOverclock to static policies
that use a single frequency.
The \textbf{Synthetic} workload simulates a server that periodically (every 100 secs)
receives a batch of compute-intensive requests and processes them as quickly as possible, then is idle until the next batch arrives.
This workload only benefits from overclocking during its request-processing phases.
Performance is measured as the total time to complete a fixed number of batches.
\textbf{ObjectStore} is a distributed key-value server running at
high load that always benefits from overclocking.
Performance is reported as P99 latency.
\textbf{DiskSpeed} is a disk-bound workload that does not benefit from overclocking.
Performance is reported as throughput in requests/sec.

Figure~\ref{fig:smart_overclock} shows the normalized peformance and power drawn by the three workloads
at various static frequency settings and when using SmartOverclock.  It shows that SmartOverclock
provides the highest or second highest performance, indicating that it is overclocking workloads when
they benefit.
Statically overclocking the Synthetic workload at 2.3GHz only provides a 13\%
performance gain over SmartOverclock, yet uses twice as much power,
demonstrating the inefficiency of static policies for dynamic cloud workloads.
ObjectStore shows similar trends.  DiskSpeed illustrates the case where SmartOverclock detects the
workload's disk-bound behavior and keeps the frequency down, except for its intentional exploration
of other frequencies.

SmartOverclock does not achieve the same performance as a static 2.3GHz frequency for CPU-bound workloads for two reasons:
(1) agent exploration intentionally sacrifices short-term benefit for long-term adaptability,
and (2) learning a model requires repeated observations to learn changes in workload dynamics.
SmartOverclock sacrifices optimal peak performance for near-optimal performance and power
usage on a wide range of workloads, achieving a higher performance/power ratio.

\myparagraph{Invalid data}
We now evaluate the impact of invalid data on SmartOverclock's model accuracy for the Synthetic workload.
In Figure~\ref{fig:oc:data_safeguard}, we vary the percentage of bad data the agent collects
by randomly returning out-of-range IPS readings to the agent a fixed percentage of the time.
Without data validation, even 5\% of invalid IPS readings causes a 17\% drop in performance,
while with data validation the workload still sees optimal performance.
Eventually, too many invalid data readings will prevent the model from making a prediction
at all and the scheduling delay safeguard will get triggered,
returning cores to the nominal frequency. 


\myparagraph{Inaccurate model}
We study the SmartOverclock model safeguard on all three workloads
in Figure~\ref{fig:model_safeguard_wrong_model} by breaking the model,
causing it to consistently select the highest frequency.
Without the model safeguard, there is nothing
to stop the agent from wasting power.
On the DiskSpeed workload, this results in a 268\% increase in power draw, whereas
the model safeguard can detect this failure and increases total power draw by only 18\%.
ObjectStore benefits from overclocking and so a broken agent that always overclocks
still achieves good results.
However, the workload could change phases at any time without
the agent changing its overclocking decision.


%

\myparagraph{Delayed predictions}
Next, we turn to the effectiveness of \system's decoupled non-blocking design in preventing
delayed predictions from impacting node health.
We study the worst case occurence of a 
delay during phase changes in the Synthetic workload,
which can cause the agent to waste power by
overclocking an idle workload.
%
%
We inject a 30-second delay in the Model thread when the workload finishes processing a batch
and compare \system's non-blocking Actuator to a blocking version that waits to change
core frequency until a prediction is available.
As Figure~\ref{fig:scheduling_safeguard} shows, the blocking agent overclocks
the workload for 30 seconds into its idle phase, increasing power consumption by 36\%.
The non-blocking agent waits a maximum of 5 seconds for a prediction from the model.
In the absence of fresh predictions, it restores the node to a safe state (nominal frequency), consuming only an additional 3\% of power.


\myparagraph{Actuator safeguard}
Fianlly, we evaluate the SmartOverclock actuator safeguard, which uses $\alpha$
to detect when the workload is in a stable phase of low CPU utilization.
Many cloud workloads include VMs that are transiently idle for many minutes at a time (\eg a VM
that runs periodic data processing jobs for 30 minutes every hour).
During these idle periods, the Actuator safeguard completely disables overclocking
to avoid wasting power.
Figure~\ref{fig:scenario_safeguard} illustrates that the safeguard can detect and disable
the agent during periods of low activity while remaining sensitive enough to quickly detect
a period of higher CPU activity and re-enabling the
agent.

\subsection{SmartHarvest}

We next evaluate our implementation of the SmartHarvest agent in \system.
Prior work~\cite{smartharvest} provides a thorough evaluation of the benefits of machine learning for CPU harvesting,
so we focus our evaluation on the additional safety provided by \system's safeguards.
\hl{When comparing the original implementation of SmartHarvest to the \system{} implementation, we find that they have a similar number of lines of code -- 1900 in the original compared to 1990 in \system{}.
  However, the version in \system{} contains the full set of safeguards required by the framework, while the original version lacks this functionality, making it more susceptible to operational issues.
  Guiding developers to ensure that their learning agents are appropriately hardened is a primary goal of \system.
Without this hardening, these issues can be difficult to detect and debug in production, hence reducing QoS and/or platform efficiency.} \yawen{providing LOC comparison for SmartHarvest}

We evaluate the SmartHarvest agent when it tries to predict the CPU utilization of a co-located primary VM.  We use either of two latency-sensitive workloads from TailBench~\cite{kasture2016tailbench} as the primary VM: \textbf{image-dnn} which performs image recognition and \textbf{moses} which does language translation.
We measure performance of both workloads as their P99 latency.

\begin{figure*}[t]
    \includegraphics[width=\linewidth]{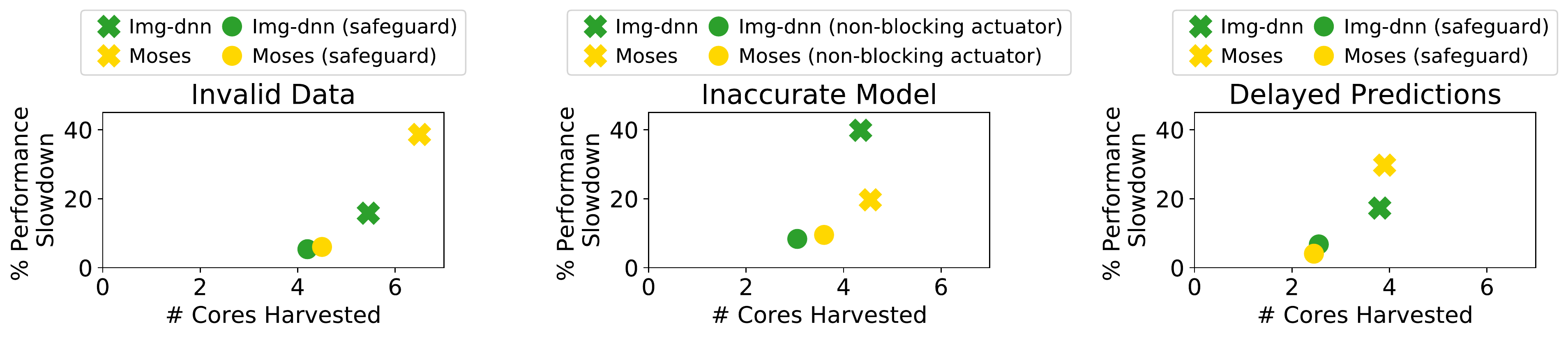}\par
    \vspace*{-.15in}
    \caption{SmartHarvest safeguards:
    The left plot shows that the data validation safeguard prevents bad data from biasing the model
    to underestimate primary VM CPU demand, reducing the impact on customer workloads by up to 4x. 
    The middle plot shows that the model safeguard reduces the impact of a broken model on workloads by up to
    4x.
    The right plot shows that the non-blocking \system implementation reduces the impact on workloads
    by up to 3x compared to a blocking agent.
    }
    \label{fig:smart_harvest_safeguard}
    \vspace*{-.1in}
\end{figure*}

\myparagraph{Invalid Data}
In the leftmost plot in Figure~\ref{fig:smart_harvest_safeguard}, we evaluate the SmartHarvest data validation safeguard,
which discards observations when the primary VM is using all available cores.
Without this safeguard, SmartHarvest consistently underpredicts the primary VM's CPU
utilization in both workloads,
causing the primary VM's P99 latency to increase by as much as 40\%.
With the safeguard, the impact on the primary VM's P99 latency is substantially less than 10\% (the acceptable performance envelope in~\cite{smartharvest}).

\myparagraph{Inaccurate model}
In the case of a broken model (middle plot of Figure~\ref{fig:smart_harvest_safeguard}),
the SmartHarvest model safeguard detects that the model is consistently underestimating the primary
VM's CPU demand.
When the safeguard is triggered, \system switches to the default predictions which alleviate
the impact on the primary VM's workload at the cost of harvesting fewer cores.

\myparagraph{Delayed predictions}
We see a similar impact on workload latency in the rightmost plot of Figure~\ref{fig:smart_harvest_safeguard},
when we insert 1-second scheduling delays during periods when the primary VM increases CPU utilization.
This worst-case scenario illustrates the importance of \system's non-blocking design.
If the agent blocks on a prediction from the model, a 1-second delay
can cause up to a 30\% increase in workload latency.
This increase happens because during the delay, the primary VM's CPU utilization
increases and it needs more cores, but the agent is blocked and cannot respond.
The non-blocking agent has no information during the delay either,
but it can quickly take the safe action of restoring all cores back to the primary VM.

\subsection{SmartMemory}

The SmartMemory agent handles delayed predictions and invalid data similarly to the other agents.
Hence, we focus on (1) demonstrating the effectiveness of adaptive access
bit scanning in reducing access bit resets and (2) the importance of \system's safeguards in protecting
workloads from too many slow tier-2 memory accesses.

\begin{figure}[t]
  \centering
    \includegraphics[width=1\linewidth]{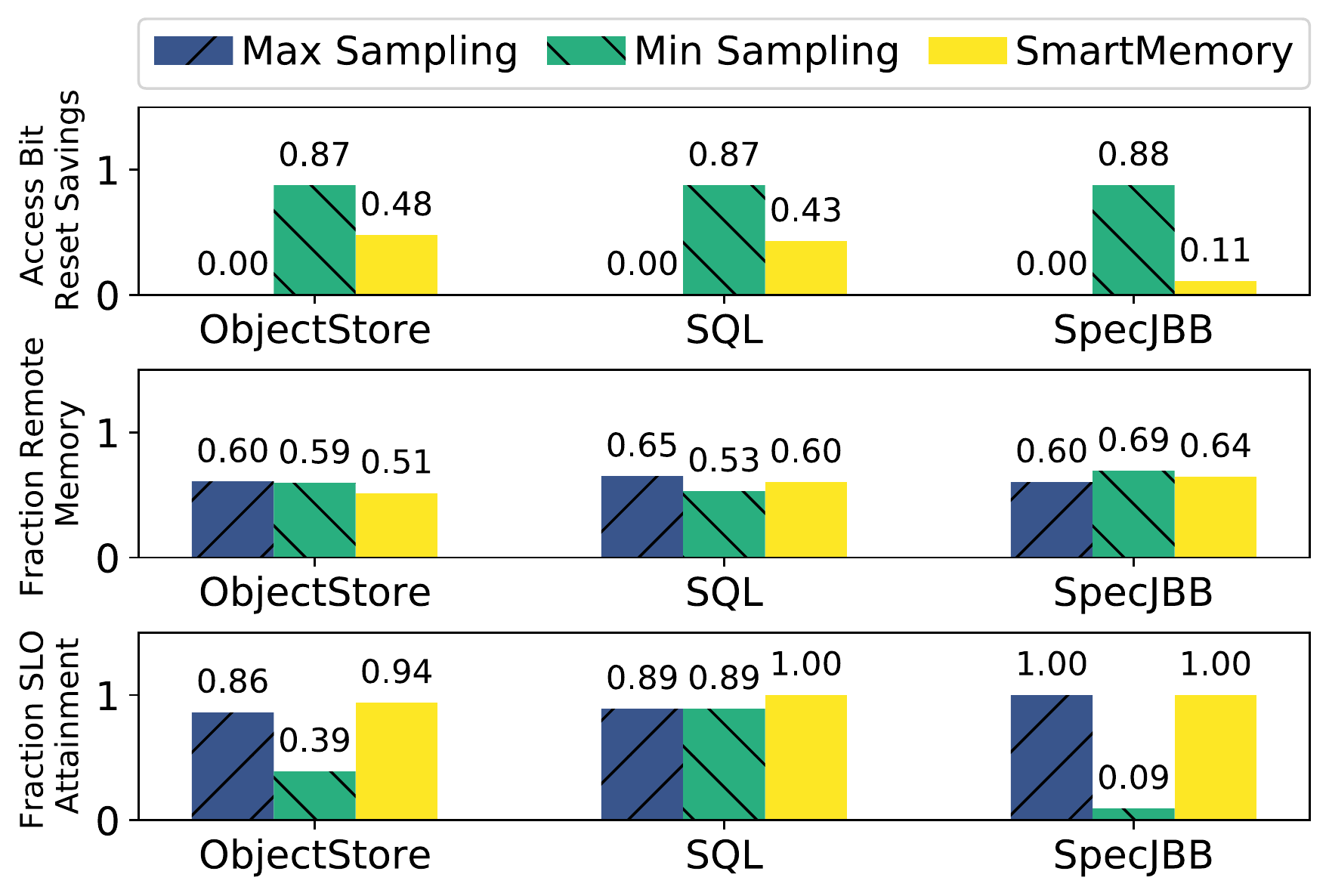}
    \vspace{-.3in}
    \caption{SmartMemory vs static access bit scanning.}
\label{fig:smart_mem_baselines}
\vspace{-.2in}
\end{figure}

In Figure~\ref{fig:smart_mem_baselines}, we compare the SmartMemory agent to
two baselines without any safeguards: always scanning at the maximum frequency (300ms)
and always scanning at the minimum frequency (9.6s).
We evaluate on three workloads: \textbf{ObjectStore},
\textbf{SQL} (a standard 
OLTP benchmark executed on SQL Server),
and \textbf{SpecJBB} (which executes
SPECjbb2000 ~\cite{specjbb} for performance evaluation of server-side Java). 
For all workloads, the agent tries to maximize remote
(tier-2) memory usage while
ensuring that at least 80\% of memory accesses are local as the service-level objective (SLO).

The top plot shows the reduction in access bit resets compared to the fastest frequency.
SmartMemory reduces access bit scans by up to a 48\%, while still
reducing local memory size by 51\% to 64\% (middle plot).
In the bottom plot, we observe the importance of \system's safeguards.
Access bit scans reflect only the current memory access patterns regardless of scanning
frequency.
Even when scanning at the maximum frequency, if the workload access patterns change,
safeguards are needed to quickly mitigate SLO violations.
Further, the bottom plot shows that sampling at the minimum frequency does not provide
enough resolution to identify the hottest batches when targeting the 80\% local accesses SLO,
resulting in SLO attainment as low as 9\%.

\begin{figure}[t]
  \centering
    \includegraphics[width=1\linewidth]{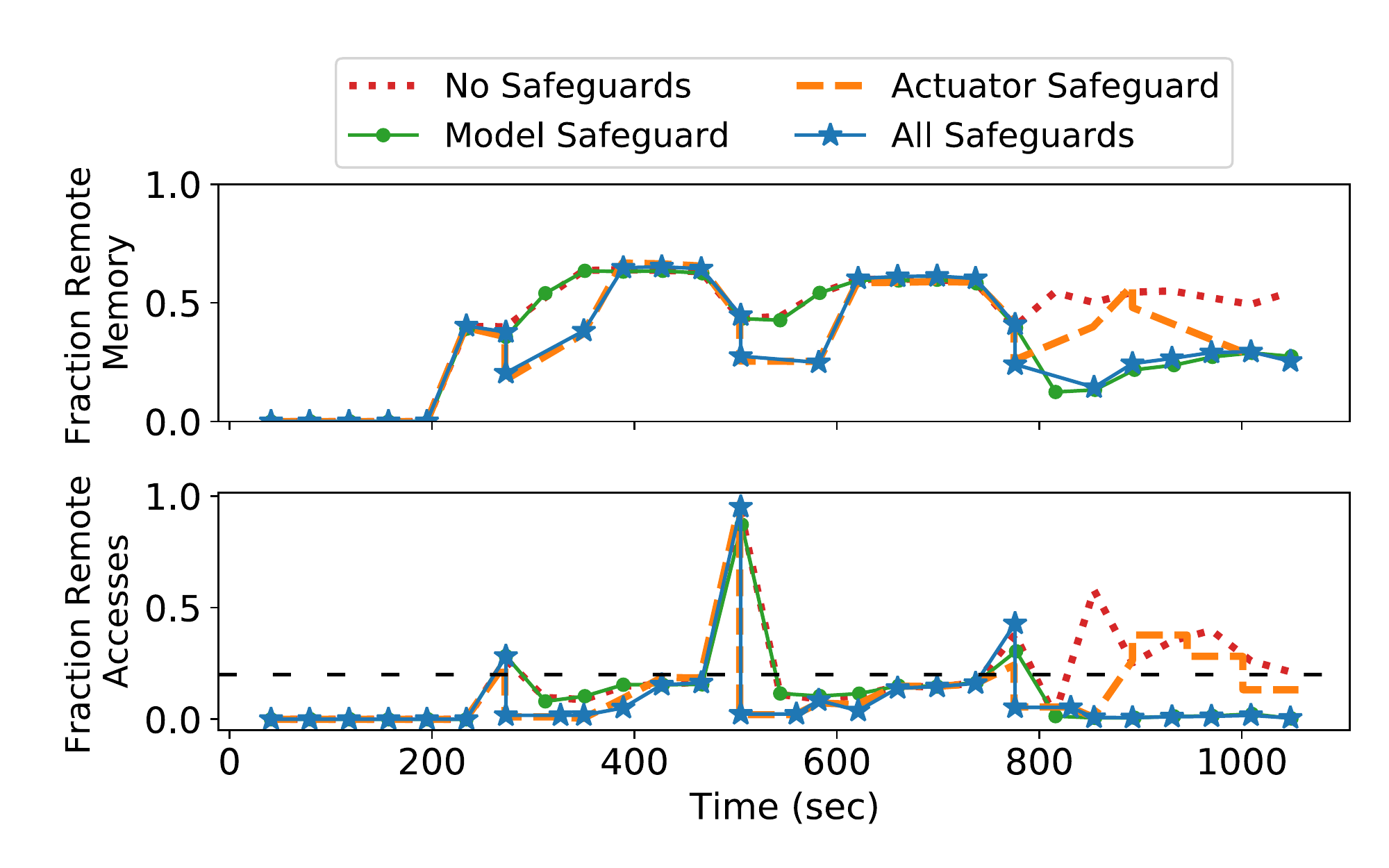}
    \vspace{-.3in}
    \caption{SmartMemory Model and Actuator safeguards.}
\label{fig:smart_mem_safeguards}
\vspace{-.2in}
\end{figure}

\myparagraph{Model and Actuator safeguards}
Figure~\ref{fig:smart_mem_safeguards} presents a more detailed evaluation of the SmartMemory
Model and Actuator safeguards.
We designed a workload that is difficult for SmartMemory to learn well:
it oscillates between running SpecJBB for 150 seconds and sleeping for 80 seconds, resulting
in frequent and rapid shifts in memory access patterns.

Without any safeguards, the SmartMemory agent only meets the SLO 66\% of the time.
When we add the Actuator safeguard, the agent can recover from instantaneous SLO violations
immediately instead of waiting for the next learning epoch.
We observe this effect at 250 seconds and 500 seconds, where the Actuator safeguard line on the bottom
plot immediately drops back below 20\% remote accesses.
However, starting around 800 seconds, the models have consistently low accuracy for several epochs and we see that
the Actuator safeguard takes multiple minutes to fully mitigate the SLO violation.
In contrast, with the Model safeguard enabled, the agent is prevented from using the inaccurate
predictions starting at 800 seconds, using the default predictions instead and avoiding the SLO violation in the
first place.

Only SmartMemory with all safeguards enabled can both avoid using inaccurate predictions
in the first place (Model safeguard), and quickly recover from SLO violations when they happen
(Actuator safeguard).
With all safeguards, SmartMemory meets the SLO 90\% of the time, even on this intentionally
difficult workload.

\section{Related Work}

We are not aware of any prior characterizations of node agents in
public cloud platforms or work on general and extensible
frameworks for implementing safe and robust on-node learning agents.

\myparagraph{Infrastructure for ML deployment}
Centralized ML systems, where models are trained offline and served online,
have become the standard deployment strategy~\cite{tf-serving,clipper,tfx,pytorch-serve, quasar, crankshaw2018prediction,velox,resourcecentral}.
Though useful for many scenarios, these systems cannot be used for the on-node learning tasks \system addresses.

\myparagraph{On-node ML} Recent works explored online learning for
improving on-node resource efficiency or workload
performance~\cite{linnos, hipster, Addanki2018PlacetoEP, mishra2017esp, fbspiral, smartchoices}.  Though effective for their particular use-cases,
they did not propose general frameworks for implementing agents or
address the deployment constraints of public cloud platforms (\eg
the need to learn at the platform level from opaque VMs, instead of
inside VMs or with application changes).  \system helps developers
build agents that run safely outside of customer VMs
without any visibility into or changes to them.

\myparagraph{Safeguards for learning} There has been some exploration
of the safety challenges involved in online ML~\cite{mao2019towards,
  surbatovich2021automatically, maeng2020adaptive, smartharvest}.  For
example, the authors of~\cite{mao2019towards} discuss a fallback policy when the model performs badly.  SmartHarvest~\cite{smartharvest} focused on protecting
the performance of customer workloads from poor predictions.  None of
these works helps developers with which issues to manage
or how to build agents in a safe and robust manner.

\section{Conclusion}

This paper explored the challenges in improving production public cloud
platforms by infusing online machine learning into their node agents.
We first surveyed the existing (non-learning) agents in \cloudplatform
and found that 35\% of the 77 agents have the potential to benefit from
learning.
We then presented \system, a general and extensible framework for developing
on-node learning agents that can operate safely under various realistic
issues, including bad data, scheduling delays, inaccurate models, and external
interference.
To demonstrate \system, we implemented three agents using it and experimentally showed
(1) the benefits of infusing learning into the agents, and
(2) how the design of \system ensures that they are robust to
a variety of failure conditions.

\bibliographystyle{ACM-Reference-Format}
\bibliography{references}

\end{document}